\documentclass{jetpl}
\twocolumn

\lat

\newcommand{\n}{\nonumber}
\def\ktf {$k_t$-factorization }

\def\a {\varepsilon}
\def\g {\gamma}

\def\F {{\cal F}}
\def\A {{\cal A}}
\def\p {{\cal P}}
\def\J {$J/\psi$ }

\def\U {$\Upsilon$ }
\def\Y {\Upsilon}
\def\u { \Upsilon  }

\def\qq {$Q\bar{Q}$ }
\def\cpc#1#2#3  {{Computer\ Phys.\ Comm.\ }  {\bf#1}, #2 (#3)}
\def\err#1#2#3  {{\it Erratum }              {\bf#1}, #2 (#3)}
\def\epjc#1#2#3 {{Eur. Phys. J. C }          {\bf#1}, #2 (#3)}
\def\dum#1#2#3  {{~}                         {\bf#1}, #2 (#3)}
\def\ib#1#2#3   {{\it ibid. }                {\bf#1}, #2 (#3)}
\def\jcp#1#2#3  {{J.\ Comp.\ Phys.\ }        {\bf#1}, #2 (#3)}
\def\jhep#1#2#3 {{JHEP }                     {\bf#1}, #2 (#3)}
\def\ijmp#1#2#3 {{Int.\ J.\ Mod.\ Phys.\ }   {\bf#1}, #2 (#3)}
\def\jpg#1#2#3  {{J.\ Phys.\ G }             {\bf#1}, #2 (#3)}
\def\mpl#1#2#3  {{Mod.\ Phys.\ Lett.\ }      {\bf#1}, #2 (#3)}
\def\ncim#1#2#3 {{Nuovo Cimento }            {\bf#1}, #2 (#3)}
\def\np#1#2#3   {{Nucl.\ Phys.\ }            {\bf#1}, #2 (#3)}
\def\npb#1#2#3  {{Nucl.\ Phys.\ B}           {\bf#1}, #2 (#3)}
\def\pan#1#2#3  {{Phys.\ At.\ Nuclei }       {\bf#1}, #2 (#3)}
\def\plb#1#2#3  {{Phys.\ Lett.\ B }          {\bf#1}, #2 (#3)}
\def\prep#1#2#3 {{Phys.\ Rep.\ }             {\bf#1}, #2 (#3)}
\def\prd#1#2#3  {{Phys.\ Rev.\ D }           {\bf#1}, #2 (#3)}
\def\prl#1#2#3  {{Phys.\ Rev.\ Lett.\ }      {\bf#1}, #2 (#3)}
\def\ptp#1#2#3  {{Prog.\ Theor.\ Phys.\ }    {\bf#1}, #2 (#3)}
\def\ppnp#1#2#3 {{Prog.\ Part.\ Nucl.\ Phys.\ } {\bf#1}, #2 (#3)}
\def\ps#1#2#3   {{Physica Scripta }          {\bf#1}, #2 (#3)}
\def\rmp#1#2#3  {{Rev.\ Mod.\ Phys.\ }       {\bf#1}, #2 (#3)}
\def\rpp#1#2#3  {{Rep.\ Prog.\ Phys.\ }      {\bf#1}, #2 (#3)}
\def\sa#1#2#3   {{Sci. Acta}                 {\bf#1}, #2 (#3)}
\def\sjnp#1#2#3 {{Sov.\ J.\ Nucl.\ Phys.\ }  {\bf#1}, #2 (#3)}
\def\spj#1#2#3  {{Sov.\ Phys.\ JETP }        {\bf#1}, #2 (#3)}
\def\spjl#1#2#3 {{Sov.\ JETP Lett.\ }        {\bf#1}, #2 (#3)}
\def\spu#1#2#3  {{Sov.\ Phys.-Usp.\ }        {\bf#1}, #2 (#3)}
\def\yaf#1#2#3  {{Yad.\ Fiz.\ }              {\bf#1}, #2 (#3)}
\def\zp#1#2#3   {{Zeit.\ Phys.\ }            {\bf#1}, #2 (#3)}
\def\zpc#1#2#3  {{Z.\ Phys.\ C }             {\bf#1}, #2 (#3)}
\def\etal {{\it et al. }}

\title{Production and polarization of \U mesons \\
       in the \ktf approach in more detail}

\rtitle{Upsilon polarization in more detail}

\sodtitle{Production and polarization of \U mesons 
in the \ktf approach in more detail}

\author{S.\,P.\,Baranov$^+$\/\thanks{e-mail: baranov@sci.lebedev.ru},
        N.\,P.\,Zotov$^*$\/\thanks{e-mail: zotov@theory.sinp.msu.ru}}

\rauthor{S.\,P.\,Baranov, N.\,P.\,Zotov}

\sodauthor{Baranov, Zotov}

\address{%
$^+$P.N.Lebedev Institute of Physics, 119991 Moscow, Russia\\~\\
$^*$ D.V.Skobeltzyn Institute of Nuclear Physics, 119991 Moscow, Russia}

\dates{20 October 2008}{*}

\abstract{%
In the framework of the \ktf approach, the production and polarization 
of \U mesons at the Fermilab Tevatron is considered, and a comparision 
of the calculated double differential distributions and spin alignment 
parameter $\alpha$  with the D0 experimental data is shown.
We argue that measuring the double differential cross section and
the polarization of upsilonium states can serve as a crucial test 
discriminating two competing theoretical approaches to the parton 
dynamics in QCD.}

\PACS{12.38.Bx, 13.85.Ni, 14.40.Gx}

\begin{document}

\maketitle

\section{Introduction}

Nowadays, the production of heavy quarkonium states at high energies
is under intense theoretical and experimental study \cite{ref1,ref2}.
The production mechanism involves the physics of both short and long
distances, and so, appeals to both perturbative and nonperturbative
methods of QCD. This feature gives rise to
two competing theoretical approaches known in the literature as the
color-singlet \cite{BaiBer,GubKra} and color-octet \cite{ChoLei} models.
According to the color-singlet approach, the formation of a colorless
final state takes place already at the level of the hard partonic
subprocess (which includes the emission of hard gluons when necessary).
In the color-octet model, also known as nonrelativistic QCD (NRQCD),
the formation of a meson starts from a color-octet \qq pair and proceeds
via the emission of soft nonperturbative gluons.
The former model has a well defined applicability range and has already
demonstrated its predictive power in describing the \J production at
HERA, both in the collinear \cite{Kraem} and the \ktf \cite{j_dis}
approaches. As it was shown in the analysis of recent ZEUS \cite{ZEUS}
data, there is no need in the color-octet contribution, neither in the
collinear nor in the \ktf approach.
The numerical estimates of the color octet contributions extracted from 
the analysis of Tevatron data are at odds with the HERA data, especially 
as far as the inelasticity parameter $z=E_{\psi}/E_{\g}$ is concerned 
\cite{KniZwi}.
In the \ktf approach, the values of the color-octet contributions
obtained as fits of the Tevatron data appear to be substantially smaller
than the ones in the collinear scheme, or even can be neglected at all
\cite{j_tev,Teryaev,Chao1,Vasin}.

Recently, the results of new theoretical caclulations of the 
next-to-leading (NLO) and next-to-next-to-leading (NNLO) order corrections
to colour singlet (CS) quarkonium production have been obtained 
in the framework of standard pQCD~\cite{maltoni}. In the region of moderate 
$p_T$ ($p_T\ge 10$ GeV), these corrections enhance the color singlet 
production rate by one order of magnitude and even larger.
These new results are in much better agreement with the \ktf predictions
than it was seen for leading order collinear calculations.

In the present note we follow the guideline of our previous publication
\cite{BZ} and show a more detailed analysis of the production and
polarization of \U mesons at the Tevatron conditions using the \ktf
approach.

\section{Theoretical framework}

In the \ktf approach, the cross section of a physical process is calculated 
as a convolution of the off-shell partonic cross section $\hat{\sigma}$ and 
unintegrated parton distribustions ${\F}_g(x,k_{T}^2,\mu^2)$, which depend 
on both the longitudinal momentum fraction $x$ and transverse momentum 
$k_{T}$:
\begin{eqnarray}
  \sigma_{pp} &=&
  \int {\F}_g(x_1,k_{1T}^2,\mu^2)\,{\F}_g(x_2,k_{2T}^2,\mu^2)\times
  \n \\  &\times& 
   \hat{\sigma}_{gg}(x_1, x_2, k_{1T}^2, k_{2T}^2,...)
   \,dx_1\,dx_2\,dk_{1T}^2\,dk_{2T}^2.
\end{eqnarray}
In accordance with the \ktf prescriptions \cite{GLR83,Catani,Collins,BFKL},
the off-shell gluon spin density matrix is taken in the form
\begin{equation} \label{epsglu}
 \overline{\a_g^{\mu}\a_g^{*\nu}} =
  p_p^{\mu}p_p^{\nu}x_g^2/|k_{T}|^2 = k_{T}^\mu k_{T}^\nu/|k_{T}|^2.
\end{equation}
In all other respects, our calculations follow the standard Feynman
rules.

In order to estimate the degree of theoretical uncertainty connected
with the choice of unintegrated gluon density, we use two different
parametrizations, which are known to show the largest difference with
each other, namely, the ones proposed in Refs. \cite{GLR83,BFKL}
and \cite{Bluem}.

In the first case \cite{GLR83}, the unintegrated gluon density is derived
from the ordinary (collinear) density $G(x,\mu^2)$ by differentiating it
with respect to $\mu^2$ and setting $\mu^2=k_T^2$.
Here we use the leading order Gl\"uck-Reya-Vogt (LO GRV) set \cite{GRV98} 
as the input colinear density. In the following, this will be referred to 
as dGRV parametrisation.
The other unintegrated gluon density \cite{Bluem} is obtained as a solution
of leading order Balitsky-Fadin-Kuraev-Lipatov (BFKL) equation \cite{BFKL} 
in the double-logarithm approximation. Technically, it is calculated as 
a convolution of the ordinary gluon density with some universal weight 
factor. In the following, this will be referred to as JB parametrisation.

The production of $\Y(1S)$ mesons in $pp$ collisions can proceed via 
either direct gluon-gluon fusion or the production of $P$-wave states 
$\chi_b$ followed by their radiative decays $\chi_b{\to}\u{+}\g$.
The direct mechanism corresponds to the partonic subprocess
$g+g\to\u+g$
which includes the emission of an additional hard gluon in the final
state. The production of $P$-wave mesons is given by
$g+g\to\chi_b,$
and there is no emission of any additional gluons.
As we have already argued in our previous publication \cite{BZ}, we see 
no need in taking the color-octet contributions into consideration.

The polarization state of a vector meson is characterized by the spin
alignment parameter $\alpha$ which is defined as a function of any
kinematic variable as
\begin{equation}\label{alpha}
 \alpha(\p)=(d\sigma/d\p -3d\sigma_L/d\p)/(d\sigma/d\p +d\sigma_L/d\p),
\end{equation}
where $\sigma$ is the reaction cross section and $\sigma_L$ is the part
of cross section corresponding to mesons with longitudinal polarization
(zero helicity state). The limiting values $\alpha=1$ and $\alpha=-1$
refer to the totally transverse and totally longitudinal polarizations.
We will be interested in the behavior of $\alpha$ as a function of the
\U transverse momentum: $\p\equiv |{\mathbf p}_{T}|$.
The experimental definition of $\alpha$ is based on measuring the
angular distributions of the decay leptons
\begin{equation}\label{dgamma}
d\Gamma(\u{\to}\mu^+\mu^-)/d\cos\theta\sim 1+\alpha\cos^2\theta,
\end{equation}
where $\theta$ is the polar angle of the final state muon measured in
the decaying meson rest frame.

The definition of helicity and, consequently, the definition of $\alpha$
is frame-dependent. There are four commonly used different definitions of
the helicity frame: these are the recoil, the target, the Collins-Soper,
and the Gottfried-Jackson systems. In our analysis, we will basically use
the recoil system (which, at the Tevatron conditions, is the same as the
laboratory or proton-proton center-of-mass system), unless a different
choice is explicitly declared.

When considering the polarization properties of $\Y(1S)$ mesons 
originating from radiative decays of $P$-wave states, we rely upon the 
dominance of electricdipole $E1$ transitions
\footnote{In our previous paper \cite{BZ}, two somewhat different 
models were used for this process.}.
The corresponding invariant amplitudes can be written as \cite{CWT}
\begin{eqnarray}
i\A (\chi_1\to\Y\g)&\propto& \epsilon^{\mu\nu\alpha\beta}
  k_{\mu} \a^{(\chi_1)}_{\nu} \a^{(\Y)}_{\alpha} \a^{(\g)}_{\beta},\\
i\A (\chi_2\to\Y\g)&\propto&
  p^{\mu} \a^{\alpha\beta}_{(\chi_2)} \a^{(\Y)}_{\alpha}
  \left[ k_{\mu}\a^{(\g)}_{\beta}{-}k_{\beta}\a^{(\g)}_{\mu} \right],
\end{eqnarray}
with $p$ and $k$ being the momenta of the decaying meson and the emitted
photon; $\a^{(\chi_1)}$, $\a^{(\chi_2)}$, $\a^{(\Y)}$, and $\a^{(\g)}$ the 
respective polarization vectors; and $\epsilon^{\mu\nu\alpha\beta}$ the
antisymmetric Levita-Civita tensor. This leads to the following 
relations between the production cross sections for different helicity 
states (see Eq. (14) in \cite{CWT}):
\begin{eqnarray}
\sigma_{\Y(h{=}0)} &=& B_(\chi_1{\to}\Y\g)
  \left[ {\frac{1}{2}}\,\sigma_{\chi_1(|h|{=}1)}\right] \n \\
                   &+& B_(\chi_2{\to}\Y\g)
  \left[ {\frac{2}{3}}\,\sigma_{\chi_2(h{=}0)}
       + {\frac{1}{2}}\,\sigma_{\chi_2(|h|{=}1)} \right] \n \\
\sigma_{\Y(|h|{=}1)} &=& B_(\chi_1{\to}\Y\g)
  \left[            \sigma_{\chi_1(h{=}0)}
        +{\frac{1}{2}}\,\sigma_{\chi_1(|h|{=}1)}\right] \n \\
                   &+& B_(\chi_2{\to}\Y\g)
  \left[ {\frac{1}{3}}\,\sigma_{\chi_2(h{=}0)}
       + {\frac{1}{2}}\,\sigma_{\chi_2(|h|{=}1)} \right.\n \\
    && \left. \mbox{\hspace*{2cm}} + \sigma_{\chi_2(|h|{=}1)} \right]
\end{eqnarray}
The dominance of electric dipole transitions (at least for the charmonium
family) is supported by the recent experimental data collected by the
E835 Collaboration \cite{E835} at the Fermilab.

All the other essential parameters were taken as in our previous paper:
the $b$-quark mass $m_b=m_\u/2=4.75$ GeV;
the \U meson wave function $|\Psi_{\u}(0)|^2=0.4$ GeV$^3$
(known from the leptonic decay width $\Gamma_{l^+l^-}$ \cite{PDG});
the wave function of $P$-wave states $|\Psi_{\chi}'(0)|^2=0.12$ GeV$^5$
(taken from the potential model \cite{EicQui});
the radiative decay branchings
$Br(\chi_{b,J}{\to}\u\g)$ = 0.06, 0.35, 0.22 for $(J=0,1,2)$ \cite{PDG};
the renormalization and factorization scale 
$\mu_R^2 = \mu_F^2 = \mu^2 = m_{\u}^2+p_{T}^2$.

\section{Numerical results}

The results of our calculations are presented in Figs. 1-4.
Fig. 1 displays the $p_T$ dependence of the differential cross section
and spin alignement parameter $\alpha$ for four different intervals of
rapidity. Complementary to Fig. 1, Fig. 2 exhibits the rapidity dependence
of the cross section and parameter alpha for three different intervals of
$p_T$. Everywhere, we separately show the contribution from the direct
production mechanism taken solely (thin curves) and after having the 
$\chi$ decays added (thick curves). When possible, we compare our 
theoretical predictions with experimental measurements 
\cite{D0_pt}-\cite{Kuzmin}.

First of all, we notice the importance of the feed-down from $\chi_b$
deccays, without which the experimental data can hardly be understood. The
calculations seem to underestimate the cross section data by approximately
a factor of 2. This can be considered as a room for higher order 
corrections and contributions from other possible subprocesses, such as 
the associated production of $\Y +b +\bar{b}$ states. The latter was shown 
to be comparable in size with the ordinary production at high $p_T$ 
\cite{Ybb}.
Any way, the disagreement by a factor of 2 must not be taken too 
seriously,as it lies within the uncertainty connected with the choice of 
factorization and renormalization scales
\footnote{As it has been argued in \cite{AG}, the proper choice should be
rather $\mu^2=(M_\Y^2 + p_T^2)/2$ than $M_\Y^2 + p_T^2$.}.
The JB gluon density leads to significantly better agreement with the data
than the dGRV density.

While the direct and indirect production mechanisms lead to more or less
similar $p_T$ and $y$ spectra, the behavior of the polarization is very
much different. This is seen in the right parts of Figs. 1 and 2, and is
vividly shown in Fig. 3. 

Our results for the direct mechanism are also applicable to the production 
of $\Upsilon(3S)$ states, with the only  exception that the overall dimuon
rate $Br_{\mu\mu}\sigma(\Y)$ is lower by an approximate factor of 4 because 
of smaller value of the wave function ($|\Psi(0)_{3S}|^2:|\Psi(0)_{1S}|^2 
\propto\Gamma_{l^+l^-}(3S):\Gamma_{l^+l^-}(1S) = 0.44:1.34$ \cite{PDG}) 
and smaller branching fraction ($2.18\%$ versus $2.48\%$ \cite{PDG}).
In this case, the absense of the feed-down from $\chi_b$ decays would 
make the experimental sample cleaner and clearer for theoretical 
analysis
\footnote{Unless there exists an unobserved $\chi_b(3P)$ state, still
below the open $B\bar{B}$ threshold, which is yet not excluded.}.

We also have to draw attention to the fact that the behavior of the spin 
alignement parameter $\alpha$ is frame depenent, as is demonstrated in 
Fig. 4. In particular, the sharp dip of $\alpha$ at $y=0$ is only 
seen in the recoil system, but not in either of the other three helicity 
systems. This property has to be not forgotten in order that the
comparison between the theoretical and experimental results be fully 
adequate.

\section{CONCLUSIONS}

We have considered the production of \U mesons in high energy $pp$
collisions in the \ktf approach and compared the predictions on the
differential cross sections and spin alignment parameter $\alpha$ 
with new D0 and CDF data. We find a more or less reasonable agreement
in all cases.

We have argued that measuring the double differential cross sections 
and, especially, the polarization of quarkonium states in extended $p_T$
and rapidity intervals can provide interesting and important 
information on their production mechanisms.

The purest probe is provided by the polarization of $\Upsilon(3S)$
mesons. In that case, the polarization is the strongest and the
predictions are free from uncertainties coming from radiative
$\chi_b$ decays.

\section*{Acknowledgments}

This work was supported by the FASI of RF (Grant No. NS-1856.2008.2), 
the RFBR foundation (Grant No. 08-02-00896-a), and DESY Directorate in 
the framework of Moscow-DESY project on MC implementation for HERA-LHC.

\newpage
\begin{figure}
\caption{%
Differential cross section and spin alignement parameter $\alpha$
as functions of the $\Y(1S)$ transverse momentum $p_T$, integrated over 
four different rapidity intervals. The panels from top to bottom:
$|y|<0.6$; $\;0.6<|y|<1.2$; $\;1.2<|y|<1.8$; $\;1.8<|y|$.
Dashed histograms, dGRV gluon density;
dash-dotted histograms, JB gluon density.
Thin lines, the direct contribution only;
thick lines, with the feed-down from $\chi_b$ states added.
Experimental points: $\bullet$ D0 \cite{D0_pt}; 
$\circ$ CDF \cite{CDF_alp}; $\ast$ D0 (preliminary) \cite{Kuzmin} .
}
\label{fig1}
\end{figure}

\begin{figure}
\caption{%
Differential cross section and spin alignement parameter $\alpha$
as functions of the $\Y(1S)$ rapidity $y$, integrated over three different
intervals of $p_T$. The panels from top to bottom:
$p_T<3$ GeV; $\;3<p_T<8$ GeV; $\;8<p_T$ GeV.
Notation of the curves is as in Fig. 1. 
Recoil system is assumed everywhere.
}
\label{fig2}
\end{figure}

\begin{figure}
\caption{%
Fraction of longitudinally polarised $\Y(1S)$ mesons 
$d\sigma(\mbox{helicity}{=}0)/d\sigma(\mbox{all helicities})$
as function of the transverse momentum $p_T$ and rapidity $y$.
Upper panel, direct subprocess; lower panel, $\chi_b$ decays solely.
}
\label{fig3}
\end{figure}

\begin{figure}
\caption{%
Rapidity dependence of the parameter $\alpha$ as seen in the different
helicity frames (sole $\chi_b$ contribution with JB gluon densities).
Dash-dotted histograms, recoil system;
dashed histograms, target system (equivalent to Gottfried-Jackson system);
dotted histograms, Collins-Soper system.
}
\label{fig4}
\end{figure}
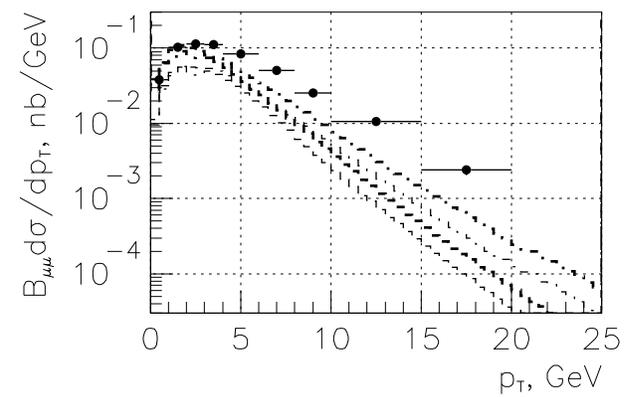
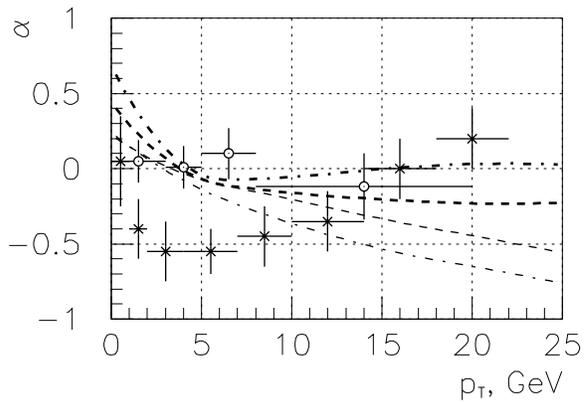
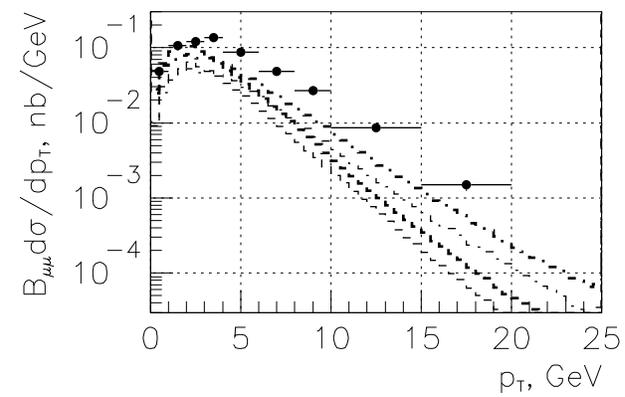
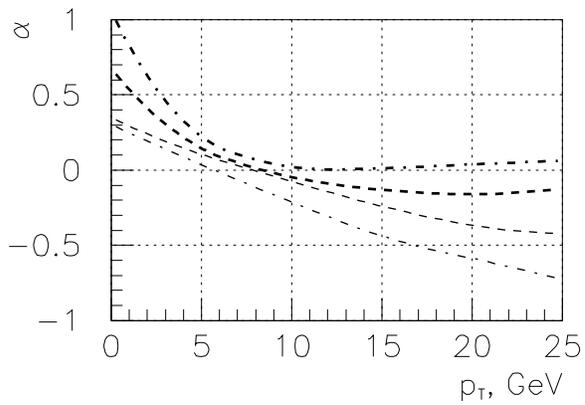
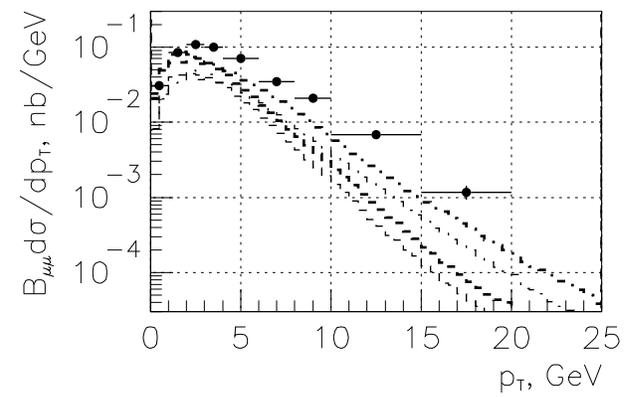
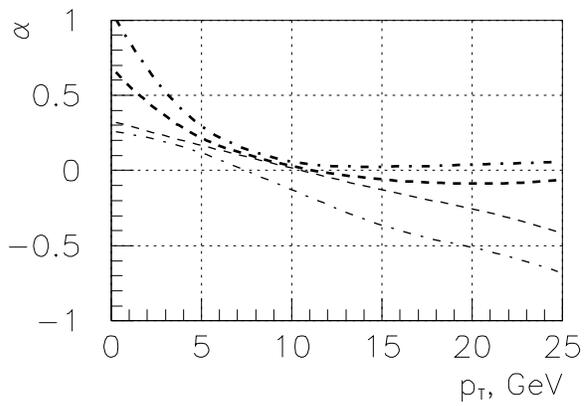
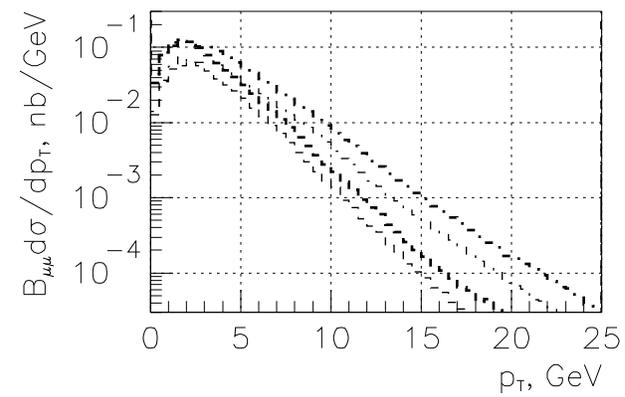
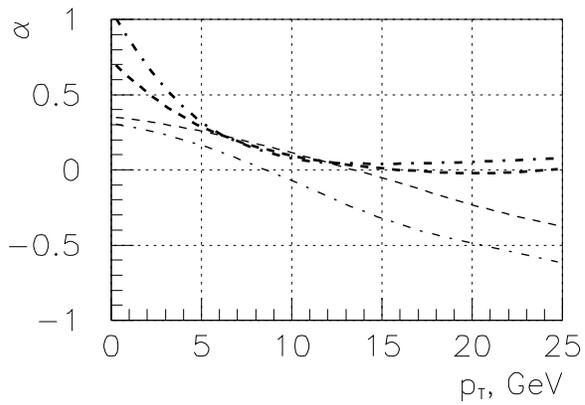
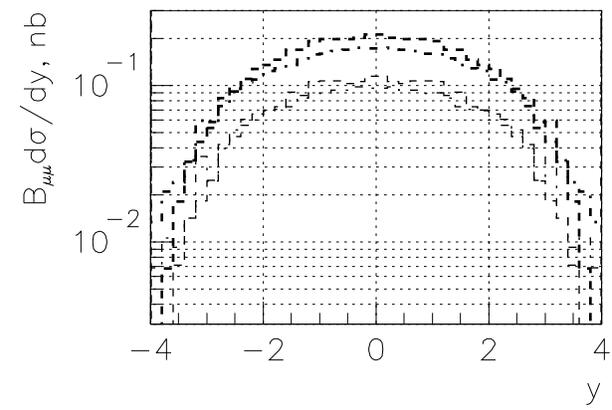
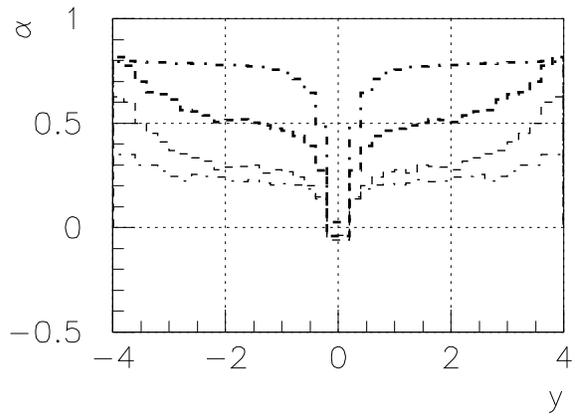
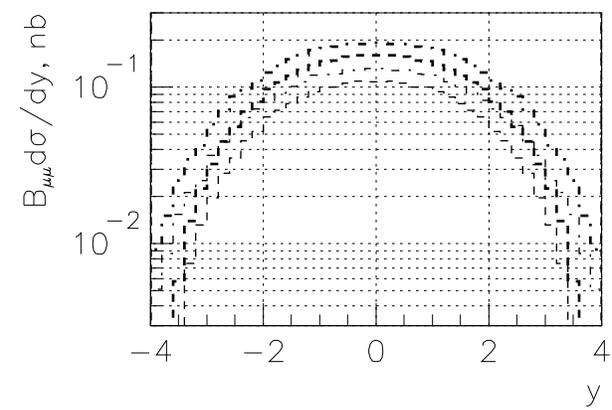
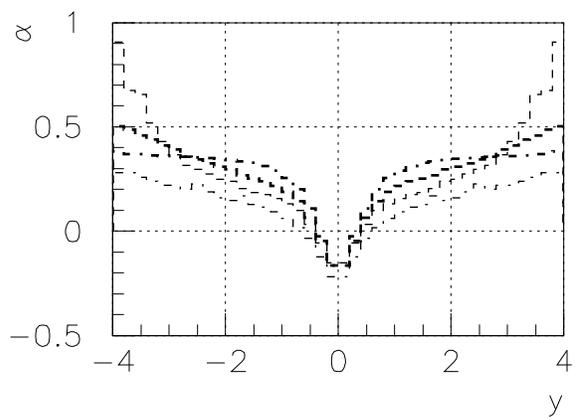
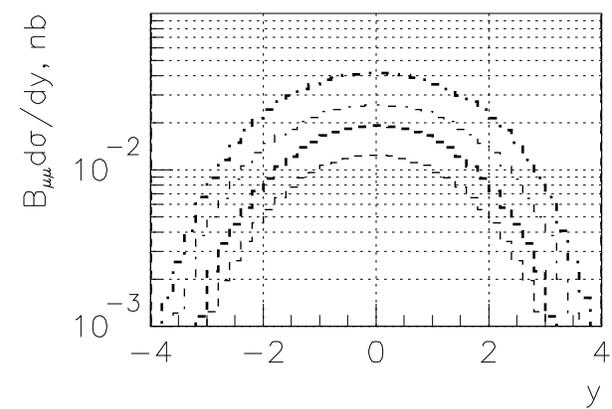
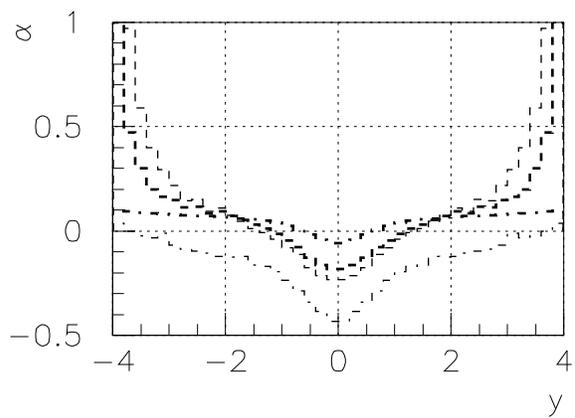
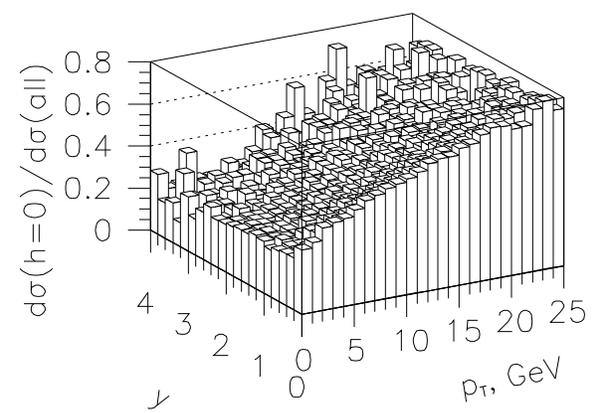
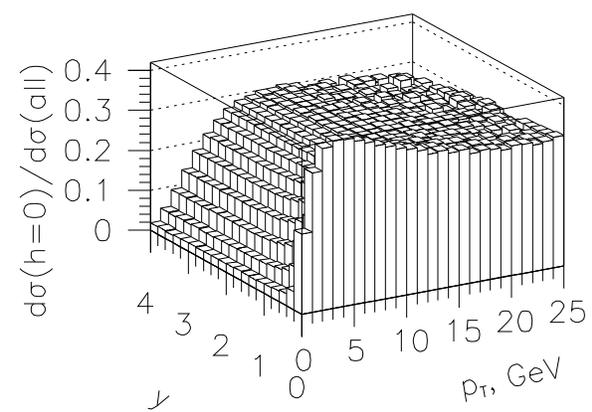
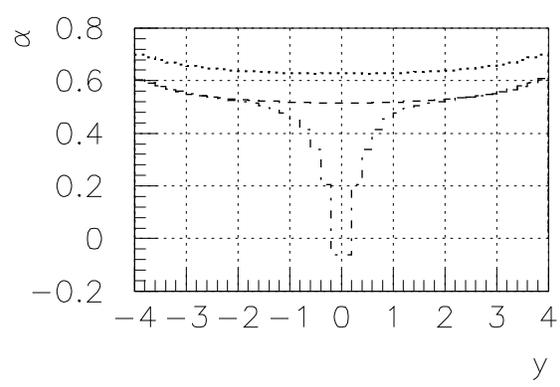
\end{document}